\documentclass[8pt]{article}  \usepackage{times}
\usepackage{graphicx}

\usepackage{color}

\begin{document}

\twocolumn[{\LARGE \textbf{The temperature dependence of lipid membrane permeability, its quantized nature, and the influence of anesthetics\\*[0.0cm]}}

{\large Andreas Blicher$^1$, Katarzyna Wodzinska$^1$, Matthias Fidorra$^{1,2}$,
Mathias Winterhalter$^3$ and Thomas Heimburg$^{1, \ast}$\\*[0.1cm]
{\small $^1$Niels Bohr Institute, University of Copenhagen, Blegdamsvej 17, 2100 Copenhagen \O, Denmark\\}
{\small $^2$Memphys, University of Southern Denmark, Campusvej 55, 5230 Odense M, Denmark\\}
{\small $^3$Jacobs University, P.O Box 750 561, 28725 Bremen, Germany\\*[0.0cm]}

{\normalsize \textbf{ABSTRACT\hspace{0.5cm} We investigate the permeability of lipid membranes for fluorescence dyes and ions. We find that permeability reaches a maximum close to the chain melting transition of the membranes. Close to transitions, fluctuations in area and compressibility are high, leading to an increased likelihood of spontaneous lipid pore formation.  Fluorescence Correlation Spectroscopy (FCS) reveals the permeability for rhodamine dyes across 100 nm vesicles. Using FCS, we find that the permeability of vesicle membranes for fluorescence dyes is within error proportional to the excess heat capacity. To estimate defect size we measure the conductance of solvent-free planar lipid bilayer.  Microscopically, we show that permeation events appear as quantized current events. Further, we demonstrate that anesthetics lead to a change in membrane permeability that can be predicted from their effect on heat capacity profiles.  Depending on temperature, the permeability can be enhanced or reduced. We demonstrate that anesthetics decrease channel conductance and ultimately lead to 'blocking' of the lipid pores in experiments performed at or above the chain melting transition. Our data suggest that the macroscopic increase in permeability close to transitions and microscopic lipid channel formation are the same physical process.}\\*[0.0cm] }}
]

\setlength{\parindent}{0cm}
\footnotesize {$^{\ast}$corresponding author, theimbu@nbi.dk --- A.Blicher and K. Wodzinska contributed equally to this work.}\\

\setlength{\parindent}{0cm}
\footnotesize{\textbf{Abbreviations:} DSC, differential scanning calorimetry; LUV, large unilamellar vesicle; }
\setlength{\parindent}{0cm}
\footnotesize{FCS, fluorescence correlation spectroscopy; BLM, black
lipid membrane;}
\footnotesize{DMPC, 1,2-dimyristoyl-sn-glycero-3-phosphocholine;}
\footnotesize{DPPC, 1,2-dipalmitoyl-sn-glycero-3-phosphocholine;}
\footnotesize{DPPG, 1,2-dipalmitoyl-sn-glycero-3-phosphocholine;}
\footnotesize{DOPC, 1,2-dioleoyl-sn-glycero-3-phosphocholine;}

\normalsize
\section*{Introduction}

Lipid membranes are regarded as insulators that are practically impermeable to ions and larger charged molecules. This is crucial for the Hodgkin-Huxley model for the action potential\cite{Hodgkin1952} but also for the interpretation of many microscopic permeation events associated with proteins. In the vicinity of the melting transition, however, membranes become permeable to water \cite{Jansen1995}, ions \cite{Antonov1980, Antonov2005} and even to large molecules. This may be of some relevance since many biological membranes are in fact found close to such a regime \cite{Heimburg2005c, Heimburg2007a}. It has been known since the early 1970s that the permeability of lipid membranes approaches a maximum in the chain melting regime of lipids in the absence of proteins \cite{Papahadjopoulos1973}, and this is also likely to be so for biological membranes in the transition regime.  The reason for this enhanced permeability is the enhanced area fluctuations that lead to a maximum in the lateral compressibility \cite{Nagle1978b, Heimburg1998}. Nagle and Scott \cite{Nagle1978b} proposed that the changes in lateral compressibility lead to a facilitation of pore formation since the increased compressibility lowers the work necessary to create a membrane defect. In the melting transition one also finds domain formation. Papahadjopoulos and coworkers \cite{Papahadjopoulos1973} suggested in their seminal paper that permeation events take place at domain boundaries.  This view has also been adopted by Mouritsen and coworkers \cite{Cruzeiro1988, Corvera1992}. At domain boundaries, fluctuations in lipid state are maximal, and this suggests that domain interfaces are especially leaky.  However, this view is not identical to that of Nagle and Scott \cite{Nagle1978b} since the overall length of domain interfaces is not identical to large fluctuations.  Unfortunately, the overall quality of permeation data for membranes so far has been low.  This is partially due to difficulties in maintaining constant temperature during optical experiments. Melting transitions of single lipid LUV display half widths of less than one degree. For a quality permeability experiment on such a system, an accuracy of at least 0.1 K is required.\\
While the measurements mentioned above refer to ensembles, i.e. dispersions of vesicles, one can also monitor permeability for ions on microscopic scales. This is done either by using black lipid membrane (BLM) spanning a small hole in a teflon film with a diameter of 50-100 $\mu$m, (Montal-M\"uller technique, \cite{Montal1972}) or in patch clamp measurements where one uses membrane patches of the order of 1 $\mu$m$^2$. Surprisingly, measurements of the conductance of pure synthetic lipid membranes can display quantized steps very similar to those reported for ion channel proteins. While there is, to our knowledge, no convincing explanation of why such currents should be quantized, there are several publications documenting such channel-like behaviour. The first paper on this is from Yafuso et al. \cite{Yafuso1974} from 1974 on oxidized cholesterol films, which appeared two years before the famous paper by Neher and Sakmann \cite{Neher1976} on quantized currents through the acetylcholine receptor. Antonov and collaborators\cite{Antonov1980} were the first to show that these lipid membrane conduction events are strongly enhanced in the phase transition regime. Using pure synthetic lipids, they could switch permeation on and off merely by shifting the temperature by a few degrees. This basically rules out the possibility that one is observing proteins or peptide impurities. Kaufmann \cite{Kaufmann1983a, Kaufmann1983b, Kaufmann1989c} showed that the occurrence of such lipid ion channels depends not only on temperature but also on other intensive variables such as lateral tension and pH (i.e., the chemical potential of protons). Along the same lines, Antonov and collaborators \cite{Antonov1985} as well as G\"ogelein \& Koepsell \cite{Gogelein1984} showed that channel-like events are influenced by calcium concentration (i.e., the chemical potential of calcium). Furthermore, they depend on voltage \cite{Antonov1990}. All of these changes are also known to influence transition temperatures in lipid membranes.  Several other publications describe quantized ion currents in pure lipid membranes in the absence of proteins \cite{Boheim1980, Yoshikawa1988, Woodbury1989, Antonov2005}. In spite of the putative relevance of such findings that are well documented and which show the coupling to thermodynamical variables convincingly, this phenomenon is not well known to a broad community. \\
It is well-established that anesthetics change the melting behavior of membranes. In particular, they lead to a lowering of the transition temperature \cite{Trudell1975, Kharakoz2001, Heimburg2007c}. Thus, it seems obvious to study the effect of anesthetics on permeability. It is an interesting historical note that Charles Ernest Overton, who first demonstrated the general action of anesthetics and its relationship to membrane solubility (Meyer-Overton rule, \cite{Overton1901}) was also the first to studied the permeability of membranes to anesthetics (Overton1895). \\

In this publication we compare ensemble permeation experiments on lipid vesicles with ion conductance measurements in planar lipid membranes. We use the FCS technique to measure the ensemble permeation rate for fluorescent dyes. In particular, we study the influence of temperature and the presence of anesthetics on permeability. Then, we determine the permeability for ions in a BLM setup. Here, we vary temperature, voltage and anesthetic concentration. The goal is to demonstrate the thermodynamic couplings that lead to changes in permeability. In particular, we show that anesthetics influence the permeability in a manner closely related to their influence on the melting of lipid membranes that can be explained by freezing-point depression.

\section*{Materials and Methods}
\textbf{Chemicals}
Decane and 1-octanol were purchased from Fluka Chemie AG (Dei\-senhofen, Germany), N-hexadecane, chloroform and methanol were obtained from Merck (Hohenbrunn, Germany), n-pentane was provided by BDH (Poole, UK) and potassium chloride from J.T. Baker Analyzed (Deventer, Holland). 1,2-dipalmitoyl-sn-glycero-3-phosphocholine (DPPC), 1,2-dipalmitoyl-sn-gly\-cero-3-phosphoglycerol (DPPG) and 1,2-dioleoyl-sn-glycero-3- phosphocho\-line (DOPC) were purcha\-sed from Avanti Polar Lipids (Birmingham, AL) and used without further purification. Rhodamine 6G and tetramethylrhodamine dextrane were purchased from Invitrogen/Molecular \-Probes (Carlsbad, Ca). For all experiments, MilliQ water (18.1 M$\Omega$) was used.\\
\textbf{Fluorescence correlation spectroscopy: }\\
\textbf{Sample preparation.}
We prepared unilamellar lipid vesicles (LUV) from DPPC:DPPG=95:5 mol/mol with an Avestin extruder (Avestin Europe, Mann\-heim, Germany) using a filter with a pore size of 100 nm diameter. Extrusion was performed in the presence of high concentrations of Rhodamine 6G fluorescence markers (50 $\mu$M). Subsequently, remaining free dye was removed on a G50 Sephadex column at temperatures below the melting temperature of the lipids where the membranes are nearly impermeable. We used vesicles made of a DPPC:DPPG\-=95:5 mixture. DPPG is negatively charged and was added to prevent the aggregation of vesicles. All experiments were performed in a 200mM NaCl buffer. At this salt concentration, vesicles do not aggregate but R6G (positively charged) does nevertheless not associate with the membranes. \\
\textbf{Fluorescence correlation spectrometer:}
Our inverted microscope setup has been describe in detail in \cite{Hac2005}. We used a linearly polarized continuous wave 532 nm Nd:Yag laser (Laser 2000, Wessling, Germany) with a power of 5mW. Further, we used a 1.20 NA 60$\times$ water immersion objective (Olympus; UPLAPO) and a confocal setup with pinhole sizes of 30 $\mu$m.   The fluorescence signal was detected by a SPCM-AQR-13 avalanche photo diode (Laser Components, Olching, Germany). Time scales were calibrated with a Rhodamine 6G solution at 296 K with a known diffusion coefficient of $D=3\cdot10^{-6}{\textrm{cm}}^{2}/ {\textrm{s}}$ at 22$^{\circ}$C. The signal from the APD was analyzed using a FLEX\-5000/fast correlator card by Correlator.com (Bridgewater, NJ). In order to adjust temperatures in the FCS experiment, both sample and microscope objective were temperature-controlled using a HAAKE DC30 K20 (Waltham, USA) waterbath. Additionally, the entire setup was heated by a ventilating radiator so that ambient temperature was close to the experimental temperature at the objective.\\
\textbf{Calorimetry: }
DSC experiments were performed on a VP-DSC Calorimeter (MicroCal, Nort{-}hampton, MA, USA) with a scan rate of $5^{\circ}$C/h. The calorimetric experiment related to the FCS experiment (Fig. \ref{Figure2}) was performed on the same sample as the one used in the FCS experiments.  For the DCS experiments related to the BLM experiments (Figs. \ref{Figure4} and \ref{Figure5}), the lipid samples were prepared by pre-dissolving in chloroform and drying the solvent under vacuum overnight. The dried lipid mixtures were dispersed in MilliQ water to a final concentration of 20mM.  The buffer used was the same as for the BLM experiments. Before filling the calorimeter, the solutions were degassed for 10 minutes.\\
\textbf{Black lipid membranes: }
Planar bilayers were formed over a round aperture in a Teflon film of 25$\mu$ m thickness, dividing two compartments of a teflon chamber embedded in a brass block that could be heated by a circulating water bath. The aperture of $\approx$80 $\mu$m radius was punctured by a steel needle (experiments in Figs. \ref{Figure5} and \ref{Figure6}) or by an electric spark (Fig. \ref{Figure4}). Lipid mixtures were prepared from chloroform-methanol solutions. The samples were dried under a weak flow of nitrogen gas/air and placed under vacuum overnight to remove the residual solvent. The aperture in the Teflon film was prepainted with 5\% hexadecane in pentane. The BLMs were painted with DOPC:DPPC 2:1 lipid solutions in decane/chloroform/methanol 7:2:1 and formed following the method described by Montal and M\"uller \cite{Montal1972}. The two compartments of the Teflon block were filled with unbuffered 150mM KCl (pH~6.5). Lipid solution (25mg/ml) was spread on the buffer surface in each compartment (approx.\ 3 $\mu$l on each side). Ag/AgCl electrodes were placed into both compartments of the chamber. After 15-30 minutes to allow for the evaporation of the solvent, the water level of the compartments was lowered and raised several times until a bilayer was formed over the hole. The formation of a BLMs was controlled visually and by capacitance measurements (with triangular 100mV voltage input pulse). The specific capacitance of the membranes was found to be ~0.9-1.2 $\mu$F/cm$^2$. For the experiments with octanol, 15\% v/v octanol in methanol solution was prepared and added symmetrically on both sides of the Teflon chamber.\\
Conductance measurements were performed on an Axopatch 200B amplifier in voltage clamp mode connected to a DigiData 1200 digitizer (both Molecular Devices, Sunnyvale, CA, USA). Current traces were filtered with 1kHz low-pass Bessel filter and recorded with Clampex 9.2 software (Axon Instruments) on the hard drive of the computer using an AD converter with a time resolution of 0.1ms. The data was further analyzed with Clampfit 9.2 and low-pass filtered with Bessel (8-pole) filter at a cut-off frequency of 300Hz. Temperature was controlled by a HAAKE DC30 K20 (Waltham, USA) waterbath and a thermocouple (WSE, Thermocoax).\\
The BLM experiments in Fig. \ref{Figure4} were performed in the laboratory of M. Winterhalter in Bremen, the experiments shown in Figs. \ref{Figure5} and \ref{Figure6} were performed using a very similar setup in the laboratory of T. Heimburg in Copenhagen.\\

\textbf{The FCS correlation function of vesicles with variable content of fluorescence dyes: }
The correlation function of a freely diffusing dye in a microscope focus of Gaussian cross section is given by:
\begin{equation}
G(\tau)=\frac{1}{\left<N\right>}\left[1+\frac{\tau}{\tau_D}\right]^{-1}
\left[1+\left(\frac{r_0}{z_0}\right)^2\frac{\tau}{\tau_D}\right]^{-\frac{1}{2}}
\label{eq:theory1.1}
\end{equation}
If one has several markers of similar 'brightness', the correlation functions add, and the number of diffusing particles can be deduced from the relative amplitude of the steps in the correlation function. The situation becomes more complicated if different objects with different brightnesses are present or if there is a distribution of brightnesses since the brightness enters the correlation as a square
\begin{eqnarray}
G(\tau) & = & \frac{1}{\left(\sum_j B_j\left<N_j\right>\right)^2}
\sum_i B_i^2\left<N_i\right>\left[1+\frac{\tau}{\tau_{D_i}}\right]^{-1}\nonumber \\
 &  & \times\left[1+\left(\frac{r_0}{z_0}\right)^2\frac{\tau}{\tau_{D_i}}\right]^{-\frac{1}{2}} 
\label{eq:theory1.2}
\end{eqnarray}
where the brightness of particle species $i$ is given by $B_i=\kappa_i\epsilon_i Q_i$ \cite{Krichevsky2001}. Here, $\kappa_i$ is the efficiency of the fluorescence detector, $\epsilon_i$ is the molar extinction coefficient of the fluorophore at the wavelength of excitation, and $Q_i$ is the quantum yield of the fluorophore. For fluorophores of equal brightness, the $B_i$s cancel. In the presence of a single dye species, one obtains eq. \ref{eq:theory1.1}. \\
\begin{figure*}[ht!]
    \begin{center}
	\includegraphics[width=16.5cm]{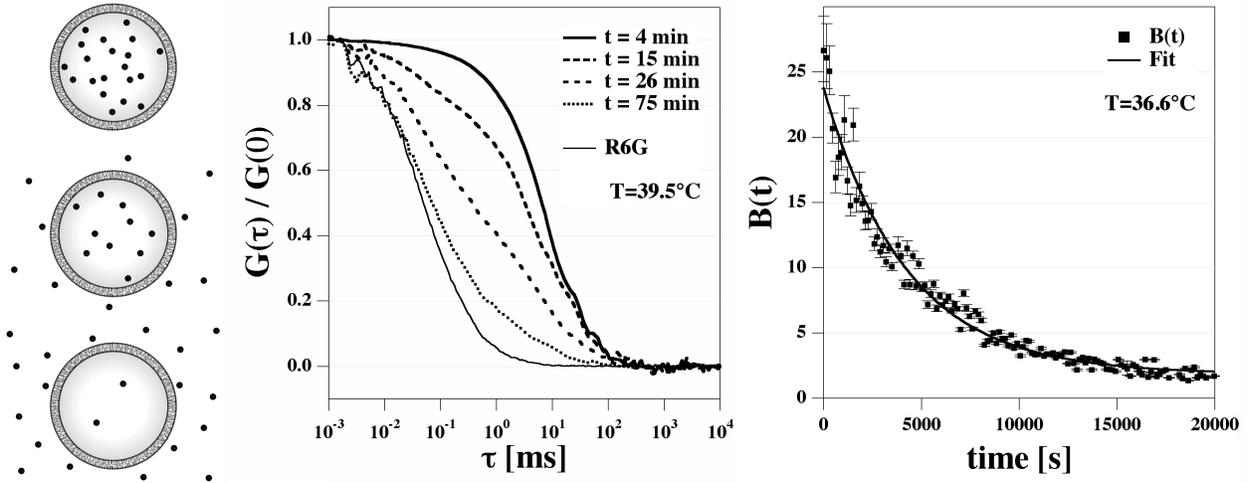}
	\parbox[c]{16cm}{\caption{\small\textit{Principle of the measurement of permeation of the fluorescence dye R6G in the FCS experiment. Left: Schematic drawing. Dye molecules trapped in vesicles smaller then microscopic focus size diffuse with the characteristic time constant of the vesicles. With time, molecules permeate through the vesicle membranes and diffuse with the characteristic time constant of free molecules. Center: The correlation profiles of this situation show two correlation steps corresponding to free molecules and vesicles. The relative step size changes as a function of time such that the number of free molecules increases and the number trapped in vesicles decreases. Five experiments are shown that were performed after the times indicated. Right: The mean number of dye molecules per vesicle (calculated from the mean brightness, see text) decreases exponentially with time. Data shown have been recorded on DPPC:DPPG (95:5 mol/mol) LUV in 200mM NaCl at temperatures of 39.5$^{\circ}$C (center panel) and 36.6$^{\circ}$C (right panel) below the chain melting transition maximum (see Fig. \ref{Figure2}). At 36.6$^{\circ}$C the time constant of permeation is 4200s. Each data point corresponds to an FCS experiment as shown in the center panel.}
	\label{Figure1}}}
    \end{center}
\end{figure*}
\textbf{Permeation through vesicles studied by FCS}
In this work we wish to monitor the permeation of fluorescence markers from the inside of a 100nm vesicle to the outside. These vesicles are smaller than the microscope focus. Even though all markers have the same brightness in this particular experiment, a vesicle that contains \textit{n} dye molecules is \textit{n} time brighter than a single dye molecule (neglecting quenching effects). In particular, vesicles do not all contain precisely the same number of fluorophores. The dye entrapped in vesicles and that free in solution is rhodamine 6G. Thus, the difference in brightness between freely diffusing single dye molecules and the vesicles is directly proportional to the number of dye molecules trapped within the vesicle. Therefore, in the following only the number of dye molecules per vesicles will appear. Not all vesicles will contain precisely the same number of molecules. From statistics, one expects a Poisson distribution of the dye content of the vesicles:
\begin{equation}
p(B_i)\approx\frac{\exp(-\widetilde{B})\widetilde{B}^i}{i!}
\label{eq:theory1.3}
\end{equation}
where $\widetilde{B}$ is the mean number of dyes per vesicle. If one inserts this distribution into eq. \ref{eq:theory1.2}, one obtains 
\begin{eqnarray}
G(\tau) & = & \frac{\left<N_{R6G}\right>}{\left(\left<N_{R6G}\right>+\widetilde{B}
\left<N_V\right>\right)^2}\left[1+\frac{\tau}{\tau_{R6G}}\right]^{-1}\times\nonumber\\
 & & \left[1+\left(\frac{r_0}{z_0}\right)^2\frac{\tau}{\tau_{R6G}}\right]^{-\frac{1}{2}} + 
\label{eq:theory1.4}
\end{eqnarray}
\begin{eqnarray}
 &  & + \frac{\widetilde{B}^2\left<N_{v}\right>\left(1+\widetilde{B}^{-1}\right)}
{\left(\left<N_{R6G}\right>+\widetilde{B}\left<N_V\right>\right)^2}\times\nonumber\\
 & & \left[1+\frac{\tau}{\tau_{V}}\right]^{-1}\left[1+\left(\frac{r_0}{z_0}\right)^2\frac{\tau}{\tau_{V}}\right]^{-\frac{1}{2}}\nonumber
\label{eq:theory1.4b}
\end{eqnarray}
The detailed derivation can be found in the M.Sc.\ thesis of A. Blicher \cite{Blicher2007}. This equation contains five unknown parameters that must be determined from experiment. The correlation function yields the two diffusion correlation times $\tau_{R6G}$ and $\tau_{V}$ of Rhodamine 6G and the vesicles, respectively. The mean number of free dye molecules in the focus, $\left<N_{R6G}\right>$, the mean number of vesicles in the focus, $\left<N_{V}\right>$, and the mean number of dye molecules per vesicle, $\widetilde{B}$, can be obtained from the absolute fluorescence intensity of the uncorrelated data, and the magnitude of the two steps in the correlation function (see Fig.\ref{Figure1}). In a time-dependent experiment, the mean number of dye molecules per vesicle will decrease. The change of $\widetilde{B}$ with time is proportional to the permeation rate through the vesicular membrane.

\section*{Theory}
\subsection*{The connection between compressibility, heat capacity and permeability}
The permeability of the membrane is maximal in the transition regime. This phenomenon is related to the well-known phenomenon of the opalescence of fluids close to critical transitions. In this regime, fluctuations in density may approach the length scale of visible light. Einstein \cite{Einstein1910} calculated the work required to generate density fluctuations on the length scale of light. In particular, he determined the work necessary to change the density of box of size $L^3$ by moving volume into another box of equal dimensions. Close to transitions this is considerably easier because the volume compressibility becomes high. In fact, one can see the increased fluctuations in liquids close to their transition in optical experiments. The problem of permeability changes and of pore formation is of a similar nature since it requires moving lipids into another region of the membrane in order to create a pore. In particular, the likelihood to form pores or holes in the membrane is closely related to fluctuations in density. The problem can now be posed as follows: How much work must be performed to move a lipid from a membrane segment of dimensions $L^2$ into another region of similar size? \\

The work necessary to compress a membrane (i.e., the Helmholtz free energy change) by an area $a$ is a quadratic function of the area that has the form
\begin{equation}
\Delta W(a)=\frac{1}{2}K_T^A\left(\frac{a}{A_0}\right)^2\cdot A_0
\label{eq:theory2.6}
\end{equation}
where $A_0$ is the total area of the membrane, $a$ is the size of a defect and $K_T^A$ is the isothermal area compression modulus (e.g., \cite{Nagle1978b}). The compressibility of a membrane is given by
\begin{equation}
\kappa_T^A=-\frac{1}{A_0}\left(\frac{\partial A}{\partial \Pi}\right)_T\equiv \frac{1}{K_T^A}
\label{eq:theory2.1}
\end{equation}
where $\kappa_T^A$ is the isothermal area compressibility.  The compressibility is also proportional to the area fluctuations \cite{Heimburg1998}
\begin{equation}
\kappa_T^A=\frac{\left<A^2\right>-\left<A\right>^2}{\left<A\right>kT}
\label{eq:theory2.2}
\end{equation}
while the heat capacity is proportional to the fluctuations in enthalpy \cite{Heimburg1998}: 
\begin{equation}
c_p=\frac{\left<H^2\right>-\left<H\right>^2}{kT^2}
\label{eq:theory2.3}
\end{equation}
The empirical finding that, close to transitions, the changes in area and the changes in enthalpy are proportional functions of temperature, i.e., $\Delta A(T)=\gamma_A\cdot \Delta H(T)$ \cite{Heimburg1998, Heimburg2000c, Dimova2000, Ebel2001}, leads to
\begin{equation}
\Delta\kappa_T^A=\frac{\gamma_A^2\;T}{A_0}\;\Delta c_p
\label{eq:theory2.4}
\end{equation}
or 
\begin{equation}
\kappa_T^A=\kappa_{T,0}^A+\frac{\gamma_A^2\;T}{A_0}\;\Delta c_p
\label{eq:theory2.5}
\end{equation}
Nagle and Scott \cite{Nagle1978b} assumed that the permeability can be expressed as a series expansion with respect to the area fluctuations that are also proportional to the lateral compressibility. The resulting permeability obtained by Nagle and Scott  is
\begin{equation}
P=a_0+a_2\kappa_T^A+\mbox{higher order terms}
\label{eq:theory2.7}
\end{equation}
where $a_0$ and $a_2$ are constants to be determined from experiment. Using expression \ref{eq:theory2.5} yields 
\begin{equation}
P=c_0+c_2\Delta c_p
\label{eq:theory2.8}
\end{equation}
where $c_0$ and $c_2$ are also constants to be determined from experiment. This is a simple expression directly coupled to the calorimetric experiment. In particular, it predicts that the change in permeability is proportional to the excess heat capacity, which is easily measured. This approach relies on area fluctuations and does not involve a molecular model for a pore. In particular, we do not consider a line tension around the pore. The experimental results below will justify this approach.\\
The results of this section also immediately imply that everything that changes heat capacity profiles will also change the permeability in a coherent and predictable manner. The similarity of this problem to critical opalescence implies that it should be possible to detect the critical nature of pore formation in optical experiments.

\begin{figure}[hb!]
    \begin{center}
	\includegraphics[width=8.5cm]{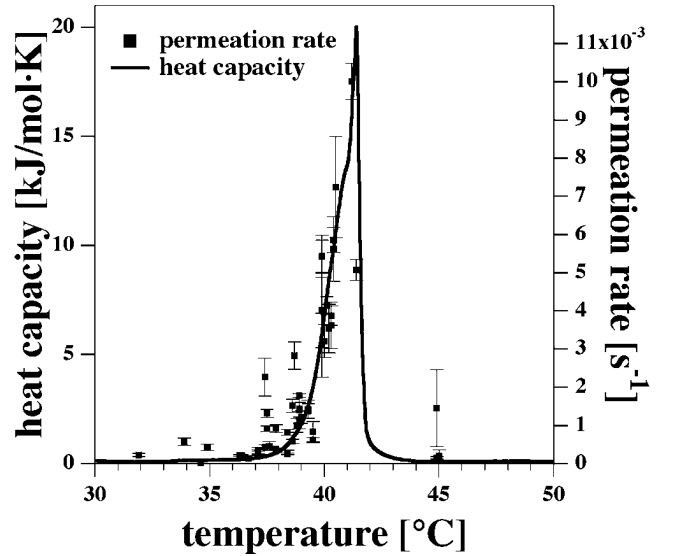}
	\parbox[c]{8cm}{ \caption{\small\textit{Permeation rate of R6G through DPPC:DPPG (95:5) LUV (200mM NaCl) as a function of temperature compared to the heat capacity profile of the identical sample. The permeation rate is closely related to the heat capacity as predicted by eq. \ref{eq:theory2.8}.}
	\label{Figure2}}}
    \end{center}
\end{figure}
\section*{Results}
We have investigated two different aspects of lipid membrane permeability: the macroscopic permeability of lipid vesicles for fluorescence markers and the permeability of planar lipid membranes for ions. While in the first setup one obtains an ensemble average over many permeation events, in the second experiment the permeation events allow us to draw conclusions regarding the possible size of the defects.  In particular, we investigate the correlation with anesthetics. We show that such drugs influence the permeability in a coherent and predictable manner.

\subsection*{Permeability of LUV measured by FCS}
Fluorescence correlation spectroscopy is a single molecule technique in which the diffusion constant and particle concentration can easily be determined on the length scale of the microscope focus. Rhodamine 6G fluorescence dyes were entrapped in lipid vesicles of 100 nm diameter. Both, vesicles and dyes are smaller than the microscope focus diameter. However, the diffusion constant of the singular dyes is much higher than that of the much larger vesicles. While the typical dwell time of a single dye molecule in our FCS setup is on the order of 100 $\mu$s, the dwell time of a fluorescing vesicle is of order 10ms. If both single dye molecules and fluorescent vesicles coexist in a solution, the correlation function will show two steps with amplitudes related to the concentrations of dyes in vesicles and free dyes.\\

Initially, all R6G molecules are trapped in the vesicles. With time, they leak out of the vesicles as shown schematically in Fig. \ref{Figure1} (left). This process can be followed in repeated FCS experiments (Fig. \ref{Figure1}, center), and the relative fraction of dye in vesicles and as free molecules can be deduced from the amplitudes of the steps in the correlation function following the procedure given in eq. \ref{eq:theory1.4} in the Materials and Methods section. As an example, we show the leakage of R6G out of vesicles made of 95 mol\% DPPC and 5 mol\% DPPG measured at 39.5$^{\circ}$C, which is about 2 K below the heat capacity maximum of the membrane melting profile.  Under these conditions permeation of dyes through the vesicular walls is quite slow. One can see two correlation steps. While the correlation times remain constant, the relative fraction of slow and fast diffusion species is changing. Fig.\ref{Figure1} (right) shows that the progress of leakage for a sample at 36.6$^{\circ}$C  ($\approx$ 5 K below the transition maximum) yields a single exponential decay with a time constant of $4800$ seconds and a permeation rate constant of 
$2.1 \cdot 10^{-4}$ s$^{-1}$. \\
\begin{figure*}[htb!]
    \begin{center}
	\includegraphics[width=16.5cm]{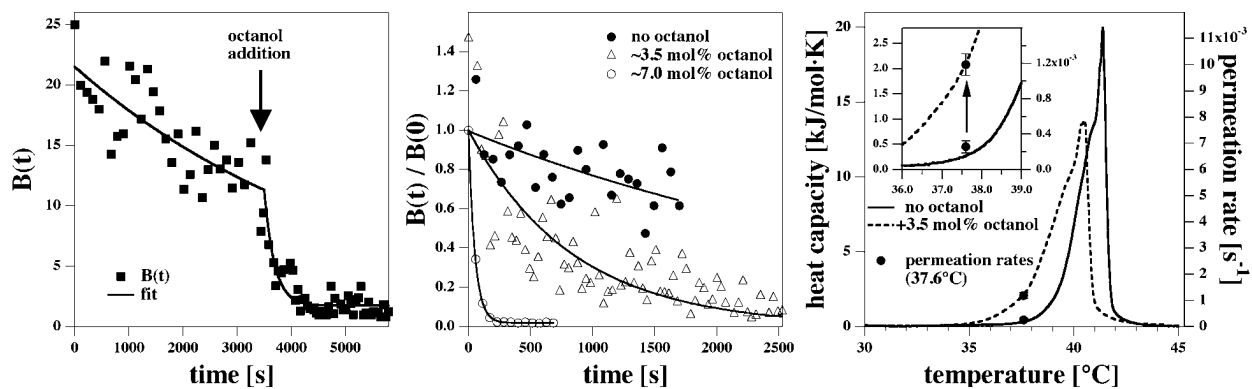}
	\parbox[c]{16cm}{ \caption{\small\textit{Effect of the anesthetic octanol on the permeation rate. Left: Permeation of R6G through DPPC:DPPG (95:5) LUV at 37.1$^{\circ}$C after the sudden addition of 1-octanol at the time indicated by the arrow. Permeation increases dramatically. Center: Permeation in the absence and the presence of 3.5mol\% and 7 mol\% octanol in the membrane at 37.6$^{\circ}$C.  Right: Heat capacity profiles of the lipid LUV in the absence of octanol (cf. Fig. \ref{Figure2}) and in the presence of 7mol\% octanol. In the presence of octanol the melting events occur at lower temperature. The filled circles indicate the permeation rates at 37.6$^{\circ}$C in the absence of octanol and in the presence of 7 mol\% octanol. The increase in permeability caused by octanol correlates nicely with the increase in heat capacity at 37.6$^{\circ}$C.}
	\label{Figure3}}}
    \end{center}
\end{figure*}

We recorded the permeation rate constants as a function of temperature and compared them to the heat capacity profile (Fig. \ref{Figure2}). It was found that the permeation rate of R6G was, within experimental accuracy, linearly related to the excess heat capacity as predicted by eq. \ref{eq:theory2.8}. In the temperature range from 36.5$^{\circ}$ to 41.4$^{\circ}$C, the permeability changed by a factor of 50. Compared to temperatures below the pretransition temperature ($< 34^{\circ}$C), the change is several orders of magnitude (data not shown). Close to the heat capacity maximum the permeation rate is about $0.01$ s$^{-1}$, which corresponds to a time constant of $100$ seconds. The dwell time of a vesicle in the microscope focus is of order $10$ ms. To obtain a correlation function of reasonable quality, one needs to record a fluorescence noise trace of at least $10$ s. Thus, the permeation rate at heat capacity maximum is close to the resolution limit of our method. Only two data points have been measured above the melting regime. These points are difficult to obtain because, during equilibration of the sample to the experimental temperature, it passes through the phase transition regime and most of the dye leaks out prior to the start of the experiment. The permeation rates for temperatures above 41.4 $^{\circ}$ C given in Fig. \ref{Figure2} represent the single exponential decay of the remaining fluorescence markers (of order 10\% of the original dye concentration). This slow decay is absent in the data at the transition point. We conclude that the permeability returns to values similar to those below the transition range, in agreement with previous studies. A control experiment with tetramethylrhodamine dextrane (MW$\approx$ 3000 g) at the melting temperature of 41.4 $^{\circ}$C  showed that this large dye remained inside of the vesicles on time scales of several hours  (data not shown). This indicates that the vesicles stay intact when passing through the melting regime and do not rupture. \\
\begin{figure*}[htb!]
    \begin{center}
	\includegraphics[width=16.5cm]{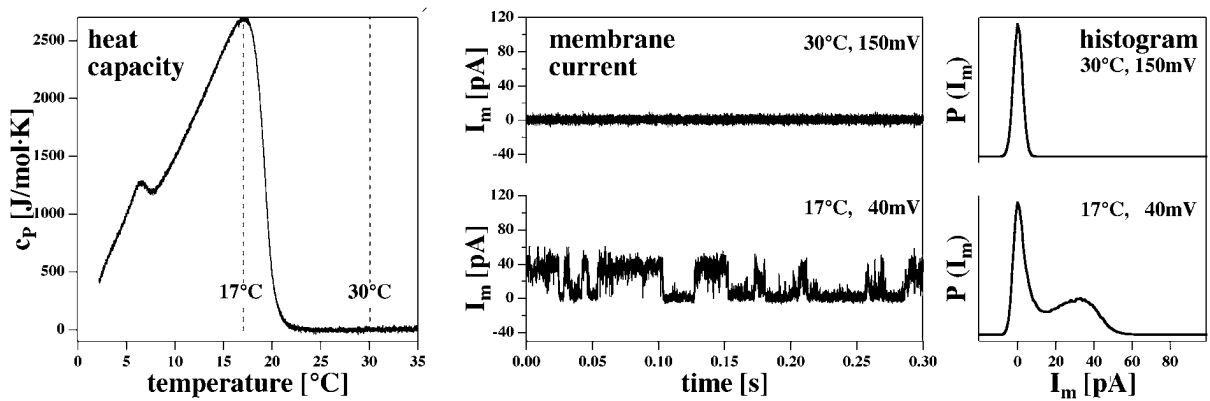}
	\parbox[c]{16cm}{ \caption{\small\textit{Spontaneous quantized currents through BLMs made of a DOPC:DPPC=3:1 mixture (150mM KCl) as a function of temperature. Left: Heat capacity traces with a heat capacity maximum around 17$^{\circ}$C. At 30$^{\circ}$C is above the melting regime in its fluid phase. Center: Current traces at 17$^{\circ}$C (40mV) and 30$^{\circ}$C (150mV). One only finds current steps in the lipid melting regime. Right: Current histograms for the two temperatures. At 17$^{\circ}$C two maxima in the histogram can be seen.  }
	\label{Figure4}}}
    \end{center}
\end{figure*}
\textbf{Influence of anesthetics:} In 2007 we showed that anesthetics act as freezing-point depressants \cite{Heimburg2007c}. At critical anesthetic dose (2.6 mol\% of anesthetics in the membrane), one obtains a melting point depression of about 0.6 degrees. Since anesthetics shift the heat capacity profile by a known value, one expects a predictable change of anesthetics on permeability. In particular, if one measures permeability below the melting temperature of the vesicles, anesthetics should increase the permeability since the melting events are moved towards experimental temperatures. If one repeats the same experiment above the melting temperature, addition of anesthetics should decrease the permeability since the transition events are moved away from experimental temperatures. This was investigated using the anesthetic 1-octanol. Fig.\ref{Figure3} (right) shows the effect of the addition of 3.5 mol\% percent of octanol (calculated from the total octanol content and a lipid/water partition coefficient of 200 \cite{Jain1978}) on the heat capacity profile. It shifts by about 0.9 K to lower temperatures, in agreement with the freezing-point depression relation derived in \cite{Heimburg2007c}. Upon addition of octanol to the vesicle dispersion at 37.6 $^{\circ}$C, the permeation rate increases dramatically (Fig.\ref{Figure3}, left. The time of octanol addition is indicated by an arrow). Dye release for different amounts of octanol is shown in (Fig.\ref{Figure3}, center). The permeation rates are $0.24\cdot 10^{-3}$s$^{-1}$ in the absence of octanol, $1.19\cdot 10^{-3}$s$^{-1}$ in the presence of 3.5 mol\% octanol and $19.8\cdot 10^{-3}$s$^{-1}$ in the presence of 7 mol\% octanol. In Fig.\ref{Figure3} (right), the permeation rate with and without octanol are compared with the heat capacity changes. At 37.6$^{\circ}$C the permeability increase caused by octanol is about 5-fold and matches exactly the change of the excess heat capacity caused by octanol at the same temperature.
\begin{figure}[htb!]
    \begin{center}
	\includegraphics[width=8.5cm]{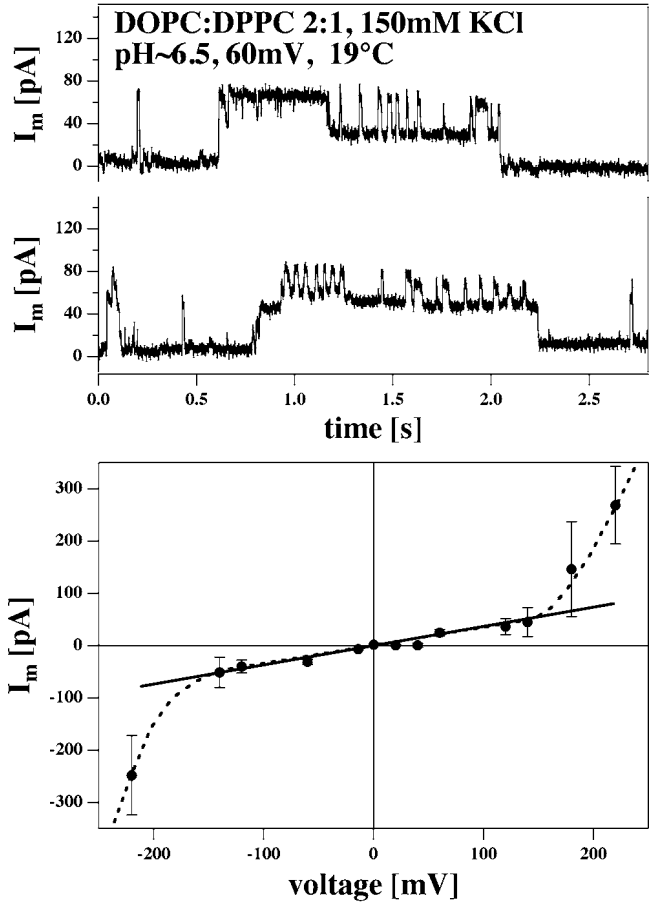}
	\parbox[c]{8cm}{ \caption{\small\textit{Currents and current-voltage relationship for  BLMs made of a DOPC-DPPC=2:1 mixture (150 mM KCl, 19$^{\circ}$C, pH$\approx$ 6.5). The membrane is in its melting regime (cf. Fig. \ref{Figure6}, left panel). Top: Current traces at a voltage of 60mV.  One can see up to five current levels with equal distance between adjacent levels. Bottom:  Current-voltage relation for the above membranes. The mean amplitude of current events is shown as a function of voltage. It is linear in the range between -150mV and +150mV. If $|$V$|$ is larger than 150 mV, the relationship is non-linear due in part to several simultaneous open events in the membrane. A total of 328 current traces of 30 second duration was used for this analysis.
}
	\label{Figure5}}}
    \end{center}
\end{figure}
\subsection*{Permeation of ions through black lipid membranes}
\begin{figure*}[htb!]
    \begin{center}
	\includegraphics[width=16.5cm]{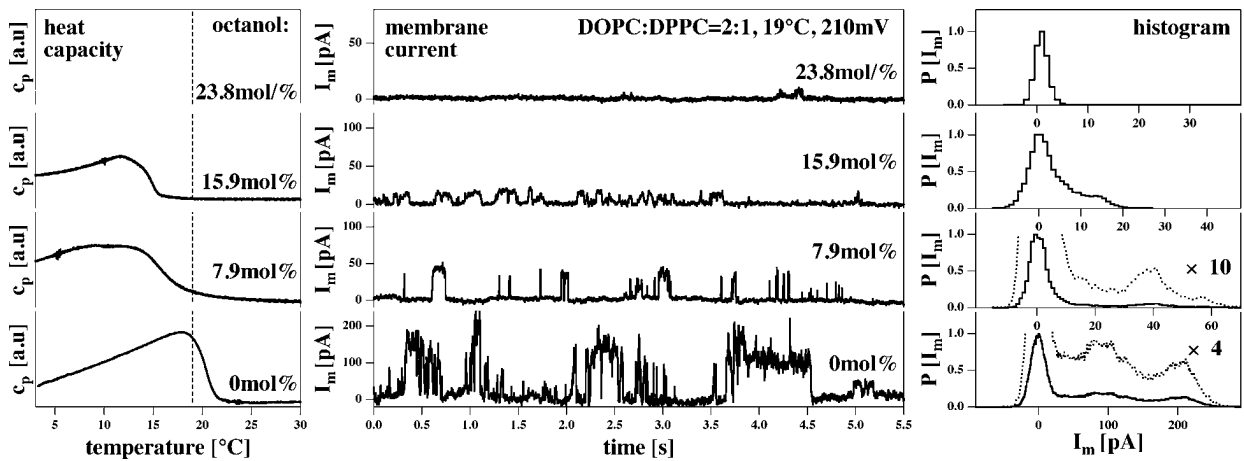}
	\parbox[c]{16cm}{ \caption{\small\textit{Influence of the anesthetic octanol on membrane currents through BLMs made of a DOPC:DPPC=2:1 mixture (150 mM KCl, 19$^{\circ}$C, pH$\approx$ 6.5) at a voltage of 210 mV. Shown are four different octanol concentrations, 0 mol\% (bottom panels), 7.9 mol\% (2nd from bottom), 15.9 mol\% (2nd from top), and 23.8 mol\% (top panels) given as mol\% of molecules in the membrane. Left: Heat capacity traces for three octanol concentrations (trace for 23.8mol\% not shown). Increasing octanol leads to a shift of the profiles towards lower temperatures. This effect is known as freezing-point depression \cite{Heimburg2007c}.  In the absence of octanol, the experimental temperature is at the c$_p$-maximum, while it is above the c$_p$-maximum in the presence of octanol. Center: Representative current traces at the four octanol concentrations. The current amplitudes and the total time in the open state become smaller with increasing octanol concentration. Right: Current histograms showing  several peaks. At higher octanol concentration, the peak positions are at lower voltages. At 23.8mol\% current events disappear completely. Note the different scales of the current axis in the center and the right panel.}
	\label{Figure6}}}
    \end{center}
\end{figure*}

It has been shown previously \cite{Antonov1980, Antonov2005} that permeation of ions through BLMs is most prominent in the phase transitions regime. Further, conductance changes occur in quantized steps similar to that reported for channel forming proteins or peptides \cite{Yafuso1974, Antonov1980, Antonov1985, Antonov2005, Gogelein1984, Kaufmann1983a, Kaufmann1983b, Kaufmann1989c}. In order to stay closer to room temperature, we used lipid mixtures in the BLM experiments which differed from those of the FCS experiments. BLMs below the transition are typically not stable. In Fig.\ref{Figure4} we show data recorded on a DOPC:DPPC=3:1 (mol/mol) mixture (150mM KCl, 1mM Tris, pH 7.4). This mixture displays a broad melting profile with a maximum at 17$^{\circ}$C (Fig. \ref{Figure4}, left panel). Above 22$^{\circ}$C the mixture is in the fluid phase. We compared current measurements at 17$^{\circ}$C and 30$^{\circ}$C (Fig. \ref{Figure4}, center panel). At 17$^{\circ}$C and a transmembrane voltage of 40mV, one sees discrete current steps with an amplitude of about 35pA (see histogram in the right hand panel) and typical channel open times of several 10ms. At 30$^{\circ}$C one does not see any current steps even at higher voltages of 150mV. This supports the notion that the transition is the regime in which such events can be found. Our observation of well-defined conductance steps indicates that this particular lipid composition is able to form well-defined defects even in the absence of peptides or proteins (e.g, \cite{Gogelein1984}). \\
Fig. \ref{Figure5} (top) shows traces from a DOPC:DPPC=2:1 (150mM KCl, 60mV, pH $\approx 6.5$, 19$^{\circ}$C). This mixture has a heat capacity maximum close to 19$^{\circ}$C.  The figure shows that one can find traces with at least four equally spaced conductance steps, probably corresponding to several lipid channels opening simultaneously. Such multiple events seem to be more frequent when one is closer to the transition regime or if the voltage is higher. The bottom panel of Fig. \ref{Figure5} shows a current-voltage relationship for the membranes used in the top panel recorded at 19$^{\circ}$C. It represents the average current for all events where I$_m$ is larger than zero. It is linear in the regime between -150mV and +150 mV. If $|$V$| >$150 mV, the overall conductance increases, probably due to a significant contribution of multiple conductance steps induced by higher voltage. Note that the electrostatic potential is an intensive thermodynamic variable that influences the state of the membrane (see discussion). From the slope in the linear regime one can calculate a channel conductance of 300$\pm$80 pS. Assuming a pore in the lipid membrane filled with electrolyte \cite{Antonov2005}, one calculates a pore radius of about 0.75nm which is practically identical to the diameter of a single lipid.  (DPPC has a cross-sectional area of 0.63nm$^2$ in the fluid state \cite{Heimburg1998} corresponding to a diameter of about 0.8nm.)  This size is similar to the radii reported for protein channels, e.g., 0.45-0.5 nm for the voltage-gated potassium channel Kv1.2 \cite{Treptow2006} or 0.3nm for the acetylcholine receptor \cite{Corry2006}. The fact that lipid ion channels display larger conductance at high voltage could be considered as a voltage gating. However, it seems likely that this effect rather reflects the influence of voltage on the transition.\\
In the section above on dye permeation through vesicles it was shown that anesthetics (octanol) influence permeation due to its influence on the heat capacity and hence the lateral compressibility. In Fig. \ref{Figure6} (left panel) the heat capacity profiles of DOPC:DPPC=2:1 membranes are shown in the absence and the presence of 1-octanol. For the DPPC:DPPG=95:1 membranes in Fig. \ref{Figure3}, the melting events are shifted towards lower temperatures. We performed BLM measurements at 19$^{\circ}$C, which is close to the transition maximum. Thus, one expects that adding octanol moves the membranes out of their transition regime and according to eq. \ref{eq:theory2.8}, the permeability should decrease. That this is in fact the case is shown in Fig. \ref{Figure6} (center and right). At 210mV, the currents in the absence of octanol show two maxima around 100 and 200pA. In the presence of 7.9 mol\% octanol the current steps range between 20 and 40pA, while in the presence of 15.9 mol\% octanol the current amplitudes are on the order of 15pA. At higher octanol concentrations (23.8 mol\%), conductance events disappear completely. This reflects the decrease in heat capacity at 19$^{\circ}$C induced by increasing octanol concentrations. Visual inspection of the current traces also seems to indicate that the mean open time per channel decreases with increasing octanol content.

\section*{Discussion}
In this paper we investigated the permeability of lipid membranes for dyes and ions using FCS and BLM measurements. In agreement with earlier studies \cite{Papahadjopoulos1973, Antonov1980, Corvera1992, Antonov2005}, we found that the permeability for both dyes and ions is maximum in the melting regime of the membranes. The FCS experiments showed that the permeation rate is, within experimental error, proportional to the heat capacity of a DPPC-DPPG=95:5 membrane. This can be understood by the large fluctuations in area near the transition. This leads to a high compressibility and, consequently, the work required to create a pore by thermal fluctuation is small (see Theory section). Several degrees below the transition the membrane was practically impermeable to dyes. Using the black lipid membrane (BLM) technique \cite{Montal1972}, we measured ion currents through lipid pores. The findings are in agreement with the dye permeation experiments. We showed that one finds quantized currents in the melting regime which are absent at higher temperatures. Due to their similarity with protein channel currents, one can consider these events as lipid ion channels. Leirer et al. \cite{Leirer2008a} showed that the mean conductance of a membrane of a lipid mixture indeed correlates well with the heat capacity (in their case of a DC$_{15}$PC:DMPC mixture). Glaser et al. \cite{Glaser1988} suggested that at sufficiently high voltages stable hydrophilic pores of 0.6-1nm can exist. This is in agreement with our findings that suggest pore sizes on the order of one lipid cross-section. The ionic radius of K$^+$ is 0.138nm, the radius of Cl$^-$ is 0.181nm and the size of the dye molecule is on the order of 1nm. In contrast to protein channels, the number of lipid channels is not fixed. They are merely fluctuating defects in membranes, and their occurrence depends on the state of the membrane as controlled by the intensive thermodynamic variables of the system (temperature, pressure, electrostatic potential, and the chemical potential of the components). We suggest that, if number and lifetime allow it, such lipid pores may fuse to form pores that are large enough to be permeable to large molecules.  If one is precisely at the transition, this may already occur at zero or small voltages. It is known that at high voltages one finds the well-documented phenomenon of electroporation \cite{Freeman1994}. At and above 500mV, long-lived pores can be produced in biological cell membranes. This effect is used to facilitate the transport of drugs into cancer cells \cite{Neumann1999}. Molecular dynamics simulations suggest that stable pores of nm-size can exist after applying transmembrane voltage pulses \cite{Tieleman2003, Boeckmann2008}. We have shown here that at voltages above 150mV the current-voltage relationship becomes non-linear so that currents are much larger. A similar finding on DC$_{15}$PC:DMPC mixtures was reported by \cite{Leirer2008a}. We further showed the existence of pores of a size sufficient to be permeable for small fluorescence dyes. One may well ask if the quantized currents found at lower voltages and electroporation found at high voltages are, in fact, the same phenomenon with the difference that the pore sizes are larger at high voltage. This matter requires further investigations. In our experiments, the BLMs typically rupture above 300mV. It should be noted here that one major disadvantage of the BLM technique is that the Teflon hole is prepainted with 5\% hexadecane or decane in pentane. Thus, the membranes contain some residual alkanes that may influence the state of the membrane to a degree that may vary between different experiments. It would be of great interest to develop the Montal-M\"uller technique so that no additional organic solvent is required --- or to repeat these studies with the patch clamp technique, as for example done in \cite{Kaufmann1989c}.\\

We have shown above that the presence of a melting transition and the accompanying area fluctuations are necessary for permeation effects. Whatever changes the state potentially effects the permeability. Here, we studied in particular the effect of anesthetics on the permeation of dyes in FCS and ions in the BLM technique. We found that in both techniques the permeability changes in agreement with the changes in heat capacity. The lipid ion channels showed smaller conductances in the presence of anesthetics. In a recent paper \cite{Heimburg2007c} we investigated the effect of anesthetics on membranes that lower the melting point of membranes. Here, we have shown that, below the melting transition, the presence of anesthetics can increase the permeability because the transition temperature is shifted towards the experimental temperature (see Fig. \ref{Figure3}). If one is above the transition, the presence of anesthetics lowers the permeability because the transition events are moved further away from experimental conditions (see Fig. \ref{Figure6}). In \cite{Heimburg2007c}, we derived the following equation for the difference in free energy between fluid and gel phase of a membrane
\begin{eqnarray}
\Delta G & = & \Delta H\;(\frac{T_m-T}{T_m}-\frac{RT}{\Delta H}x_A+\gamma_V\Delta p\frac{T}{T_m}\nonumber\\
 & & + f(\mu_{H^+}) + f(\mu_{Ca^{2+}})+f(\Psi)+...)
\label{eq:discussion1}
\end{eqnarray}
where $\Delta H$ is the transition enthalpy, $T_m$ is the melting temperature, $x_A$ is the molar fraction of anesthetics in the membrane, $\Delta p$ is the hydrostatic pressure and $\gamma_V$ is a parameter describing the relation between volume and enthalpy changes ($\gamma_V=7.8\times 10^{-10}$ m$^2$/N). This equation contains one term describing the effect of temperature, a term for the influence of anesthetics and a further term describing the effect of hydrostatic pressure. For each change in an intensive thermodynamic variable, one can add further terms into this equation, for example for pH, calcium concentration or voltage, denoted here as $f(\mu_{H^+})$,  $f(\mu_{Ca^{2+}})$,  and $f(\Psi)$. We suggested in \cite{Heimburg2007c} that whenever the value of the potential $\Delta G$ is the same, the anesthetic action is also the same. Similarly, one can state here that whenever the value of $\Delta G$ is the same, one expects the same permeability. Changes in different variables may compensate, for example that of pressure and the anesthetics concentration known as pressure reversal of anesthesia \cite{Heimburg2007c}. Therefore, it is to be expected that the decrease in permeability observed in our BLM measurements would be compensated by hydrostatic pressure.  In biological preparations that are found slightly above the transition of their membranes, one would expect an increase in conductance of lipid ion channels with increased pressure. In this context, it is interesting to note that the acetyl choline receptor displays a dependence on hydrostatic pressure, which becomes apparent at 400 bars\cite{Heinemann1987}, corresponding to a shift of the transition temperature of about 10$^{\circ}$K to higher temperatures \cite{Ebel2001}. This is the approximate difference of the melting temperature of biological membranes and physiological temperature. One would also expect that an increase in voltage would change the state of membranes because each monolayer of the membrane possesses a quite high dipole potential of order 200-500mV (e.g., \cite{Sehgal1979}). Antonov et al. \cite{Antonov1990} showed that increase in voltage increases the chain melting temperature. Here, we found a non-linear current-voltage relationship (Fig. \ref{Figure5}). The increase in conductance at higher voltages observed in our BLM measurements (Fig. \ref{Figure5}) is in agreement with Antonov's findings. However, the voltage effect is not well studied. Since the bilayer possesses two monolayers with opposing dipole potentials, the effect of voltage may be different in the two layers. Impurifications and proteins (or their chemical potentials) are further intensive variables. We have shown previously that fluctuations in lipid state are especially enhanced at lipid domain boundaries \cite{Seeger2005} and at protein interfaces \cite{Ivanova2003}. This makes domain boundaries and protein-lipid interfaces particularly susceptible for lipid ion channel formation.  \\

Previously, we have investigated the fluctuation lifetimes of lipid vesicles that are identical to the relaxation time scale after a small perturbation \cite{Grabitz2002, Seeger2007}. We found that relaxation time scales are proportional to the heat capacity. Multilamellar vesicles with very cooperative transitions display relaxation times of up to 30 s at the c$_p$ maximum. Unilamellar vesicles display less cooperative transitions with maximum relaxation times around 1-2s. If calorimetric profiles are broader (and the heat capacity is smaller), one rather expects relaxation times in the vicinity of 10-100ms. E.g., it has been estimated that the relaxation time scale in the c$_p$ maximum of lung surfactant is about 100ms. This time scale is close to what we find here for the channel life times. It has been discussed before that channel life times are in fact longer when the heat capacity is higher \cite{Seeger2007}. This has been shown in BLMs in a paper by Leirer et al. \cite{Leirer2008a} showing longer mean open times of the lipid channels close to the $c_p$-maximum of a DC$_{15}$PC:DMPC mixture that correlated with the heat capacity changes. This may explain why Antonov et al. \cite{Antonov2005} found channel lifetimes on the order of seconds at the c$_p$ maximum of DPPC membranes. As mentioned, this is the relaxation time scale for unilamellar DPPC membranes \cite{Seeger2007}. Overall, the channel lifetimes seem to be related to the fluctuation time scales of lipid membranes.\\

Finally, it is tempting to compare lipid channels with protein channels. It seems that they display very similar signatures both with respect to conductances and to mean channel open times. This is striking and raises the question of whether the physics behind these events is similar. It is remarkable that Na$^+$ channels and the acetylcholine receptor are affected by octanol in a manner similar to our lipid channels \cite{Zuo2001, Horishita2008}. The same is true for the effect of ethanol that leads to an increase in channel life times (data not shown) and which has also been reported to have a similar affect on the acetylcholine receptor \cite{Forman1999, Zuo2004}. In the literature, the effect of anesthetics on ion channels has led to the notion that anesthesia acts via a specific binding mechanism to channel proteins, a view that we have opposed in \cite{Heimburg2007c} on thermodynamic grounds. The similarity between protein channel data and lipid membrane conductance  deserves to be investigated in greater detail in future studies. At present, the precise connection between the two phenomena must remain an open question.

\section*{Conclusion}
We have shown here that lipid membranes are permeable to dyes and ions if one is close to the chain melting regime. The microscopic permeation for ions has been shown to be quantized with conductances and life times similar to those reported for proteins channels. Membrane permeation can be influenced by changes in intensive thermodynamic variables, in particular temperature, voltage and chemical potential of anesthetic molecules. Other authors have shown that permeation for ions can be influenced by calcium \cite{Gogelein1984, Antonov1985}, pH changes \cite{Kaufmann1983a, Kaufmann1983b} and changes in membrane tension \cite{Kaufmann1989c}. The same will be true for an increase of hydrostatic pressure. Thus, it must be concluded that the overall physics of lipid membrane permeation is straight forward and easily understood as a consquence of lateral area fluctuations. Since all biological membranes possess lipid membranes (and membrane proteins) it cannot be conclusively excluded that some of the quantized current events reported in the literature for proteins are in fact due to lipid channels, in particular when considering that biological membranes exist in a state close to their melting transitions \cite{Heimburg2005c, Heimburg2007a}.\\


\textbf{Acknowledgments:} Matthias Fidorra was supported by the Villum-Kann-Ras\-mussen foundation via BioNET. We acknowledge discussions with Dr. K. Kaufmann (G\"ottingen) and knowledge of earlier manuscripts that have been printed in a private edition (see reference \cite{Kaufmann1989c}). The BLM cell was build during a bachelor project of H. Krammer who was a visiting student from the LMU Munich. We further acknowledge the friendly help concerning the BLM setup by Dr. M. F. Schneider and C. Leirer (Augsburg). A paper with complementary data on the temperature dependence of lipid channels has recently been submitted by Leirer, Schneider and coworkers  \cite{Leirer2008a}. We thank Prof. Andrew D. Jackson for critical proof-reading of the manuscript.

\footnotesize{

}

\end{document}